\documentstyle[preprint,pra,aps]{revtex}
\begin{document}
\begin{center}
{\large \bf ATI as a source for multiply charged ion production
in a laser field}
\end{center}
          
\begin{center}
M.~Yu.~Kuchiev$\dagger$ 
\end{center}
\begin{center}
School of Physics, University of New South Wales\\
Sydney 2052, Australia\\
($\dagger $ on leave from A.F.Ioffe PTI, St.Petersburg, Russia)
\end{center}

\begin{abstract}
A theory for the many-electron multi-photon process
is presented. 
It is shown that after
single-electron excitation into some level in the continuum (ATI)
an inelastic collision of the excited electron with the parent atomic
particle can result in an excitation of the ion.
It may be the continuum state excitation giving
the doubly charged ion
or the discrete state which also greatly
enhances the doubly charged ion production. 
The probability of these phenomena greatly exceeds that of the 
direct ionization of a single-charged ion. The single-electron ATI 
makes possible the two-electron process even in the moderate field.
The example of two-electron excitations of He atom 
in a 780 nm laser field with intensity above $\approx 10^{14}$W/cm$^2$
is discussed.

\end{abstract}

PACS: 3280K 

\newpage

\section{Introduction}
In this paper we consider many-electron processes in a  low-frequency
laser field. One of the most exciting manifestations among these phenomena
is the creation of multiply charged ions
in a laser field. Since its  discovery (Suran and Zapesochnyi' 1975)
it has been intensively studied experimentally
for a variety of atoms and charge multiplicity of ions 
(L'Huillier $et~al$ 1982,1983a,1983b, Lompre $et~al$ 1984, 
Luk $et~al$ 1983, Boyer $et~al$ 1984, Johann $et~al$ 1986, 
Chin  $et~al$ 1985, Feldmann $et~al$ 1982a, Feldmann $et~al$ 1982b,
Agostini and Petite 1984,1985a,1985b, Dexter $et~al$ 1985, Eichmann  
$et~al$ 1987, Camus $et~al$ 1989, Zhu $et~al$ 1989,
DiMauro $et~al$ 1988, Bondar' and Suran 1993, Walker $et~al$ 1994). 
Different
experimental conditions such as a
wavelength of radiation, intensity of laser field, duration
of  laser impulse were used. 
It was recognised that  the ionization process
strongly depends on both the atomic structure
and the experimental conditions. 
Multiple ionization can proceed either as a  sequential process
or as a  collective atomic response
resulting in simultaneous removal of  several atomic electrons.
Noble gases are shown 
(L'Huillier $et~al$ 1982,1983a,1983b, Lompre $et~al$ 1984,
Walker $et~al$ 1994)
to exhibit
simultaneous, direct removal of two electrons 
below the saturation level. After saturation
the sequential ionization takes place (L'Huillier $et~al$ 1983a). 
The later statement agrees with
the result of 
(Johann $et~al$ 1986)
and with
the theoretical estimations 
(Lambropoulos 1985). 
For alkaline-earth atoms 
the two-electron mechanism  has been recently shown 
(Bondar' and Suran 1993) to be
predominant for doubly charged ion formation
in the infrared range, while in the optical
and ultraviolet ranges the sequential mechanism dominates 
(Feldmann $et~al$ 1982a, 
Agostini and Petite 1985a, Dexter $et~al$ 1985, Eichmann  
$et~al$ 1987, Camus $et~al$ 1989, Zhu $et~al$ 1989,
DiMauro $et~al$ 1988).

The resonances can in some cases
strongly manifest themselves in the formation of 
multiply charged ions.
There is strong experimental evidence showing that
creation of doubly charged alkaline-earth ions is
strongly influenced by resonances in 
neutral atoms  and single-charged ions 
(Feldmann $et~al$ 1982a, Feldmann $et~al$ 1982b,
Agostini and Petite 1984, Dexter $et~al$ 1985, Eichmann  
$et~al$ 1987, Camus $et~al$ 1989, Zhu $et~al$ 1989).
The calculation for 5-photon double-electron ionization of
Ca atom are in agreement with the resonance mechanism 
(Tang and Lambropoulos 1987).
In contrast, the resonances seem to be much less important
for ionization of noble gases where
no resonance structure was reported 
(L'Huillier $et~al$ 1982,1983a,1983b, Lompre $et~al$ 1984,
Walker $et~al$ 1994).

There were several attempts to theoretically describe
the collective mechanism of ionization in the absence of resonances.
In  (Boyer and Rhodes 1985, Sz$\ddot o$ke and Rhodes 1986)
there was considered the possibility of 
energy transfer from the outer electron shell driven by
the laser field to the inner-atomic shells. In 
(Kuchiev 1987)
 ``the antenna''  mechanism of ionization was suggested.
According to this mechanism after single-electron
ionization the ionized electron can absorb  energy
from the field and transfer it to the parent atomic
particle in the inelastic collision. Recently the idea of rescattering
of the first ionized electron on the ion
has been discussed in 
(Corkum 1993).
In  (Ostrovsky and Telnov 1987a,1987b)
there was developed 
an adiabatic theory of the multiphoton processes.
In  (Fittinghoff $et~al$ 1992)
there was considered
the possibility that the first ionized electron is taken away from the atom
so quickly that the second electron is shaken off. 
Note, however, that this idea seems to contradict
the adiabatic nature of the multiphoton ionization.  

In this paper  a theoretical approach 
to describe the direct, collective 
 many-electron process in the laser field is developed.
The basic physical idea is the following.
Suppose that first  single electron ionization takes place.
Suppose that then the ionized electron 
can undergo the inelastic collision with the parent atomic particle.
Then one can expect  the electron impact to result in the
excitation of the  ion into a discrete or continuum state.
The reason for the excitation is the fact that the ionized electron strongly 
interacts with the laser field. It can absorb and  accumulate  high energy
needed for ion excitation.
The momentum of the ionized electron oscillates due to wiggling in the
laser field. Therefore its ``energy'' is also time-dependent, it oscillates.
The time-dependent ``energy'' of the ionised electron
depends on two parameters. One of them is the time-dependent field momentum
of the electron ${\bf k}=({\bf F}/\omega)\sin \omega t$, where $F,\omega$ 
are the strength and frequency of the laser field. 
The other is the constant
translational momentum ${\bf K}$.
The total momentum is ${\bf K}+({\bf F}/\omega) \sin \omega t$. Thus the
kinetic energy of the photoelectron is $E_{\bf K}(t)=({\bf K}+
({\bf F}/\omega) \sin \omega t)^2/2$. We will show that the ion excitation due
 to electron impact is possible if the ion excitation energy $E_{\rm exc}$
is equal or less than the maximal kinetic energy of the photoelectron
\begin{equation}\label{maxE}
E_{\rm exc}={\rm max}(E_{\bf K})=
\frac{1}{2}\left ( K + \frac{F}{\omega} \right ) ^2~.
\end{equation}
This is an important inequality. It states that if the field is strong enough,
$F^2/(2\omega^2)\ge E_{\rm exc}$, 
then we can forget about the translational momentum, 
considering $K\approx 0$. For weaker fields the translational momentum 
becomes vital. Here is a point where ATI, see (Agostini $et~al$ 1979)
and  references in review (Freeman and Bucksbaum 1991), comes into play.
ATI permits the photoelectron to occupy an excited level in the continuum.
It is clear that the higher the level in the ATI spectrum is, the larger is 
the translational 
momentum.  For sufficiently high level an inequality in Eq.(1) can
be fulfilled permitting the mechanism to work. 
It is important that the necessary level in the ATI spectrum should not
be as high as the ion excitation energy is, it 
can be well below the excitation energy. Really, the energy $E_{\rm ati}$
absorbed above the
single-electron ionization threshold is equal to the averaged kinetic energy of
the photoelectron, $E_{\rm ati}=\bar E_{\bf K}=K^2/2+F^2/(4 \omega^2)$.
The latter is always less than the maximal kinetic energy, $\bar E_{\bf K}<
{\rm max}(E_{\bf K})$. The stronger the field is, the larger is the difference
between $\bar E_{\bf K}$ and ${\rm max}(E_{\bf K})$. As a result
${\rm max}(E_{\bf K})$ can exceed the ion excitation energy, while
$E_{\rm ati}$ can be below it, $E_{\rm ati}<E_{\rm exc}<{\rm max}(E_{\bf K})$.
All this means that absorption
of low energy above the single-ionization threshold
can make the considered mechanism to work resulting in high energy excitation
of the ion.
Section VII illustrates this statement 
using  an example of double ionization of He atom. For the case considered
absorption of only few (3--6) quanta above the single-ionization 
threshold is sufficient
to allow for double ionization of He which needs very high energy.
It is important  that ATI takes place
with  high probability as  the numerous experimental data presented in
(Freeman and Bucksbaum 1991)
and numerical calculations 
(Potvliege and Robin Shakeshaft 1989) demonstrate. This makes the considered 
mechanism to be very efficient.

The idea of the process considered  is illustrated by the Feynman
diagrams in Fig.1.
\\ 
\begin{picture}(200,150)(-220,-90)
\thicklines
\put(-140,0){\vector(1,0){15}}
\put(-125,0){\line(1,0){25}}
\put(-100,-2.5){\line(0,1){5}}
\put(-100,-2.5){\line(1,0){80}}
\put(-100,2.5){\line(1,0){80}}
\put(-20,-2.5){\line(0,1){5}}
\put(-20,-2.5){${\bf > }$}
\put(-150,0){$a$}
\put(-10,0){${\bf p}$}
\put(-101,1){\line(-5,2){10}}
\put(-113,7.5){\line(-5,2){10}}
\put(-125,15){\line(-5,2){10}}
\put(-60,-10){\line(0,1){7}}
\put(-60,-20){\line(0,1){7}}
\put(-60,-30){\line(0,1){7}}
\put(-60,-40){\line(0,1){7}}
\put(-140,-40){\vector(1,0){15}}
\put(-125,-40){\vector(1,0){100}}
\put(-25,-40){\vector(1,0){5}}
\put(-20,-40){\line(1,0){5}}
\put(-150,-40){$b$}
\put(-10,-40){\makebox(0,0){$c$}}
\put(-80,-65){\makebox(0,0){\bf a}}
\put(45,0){\vector(1,0){15}}
\put(60,0){\line(1,0){25}}
\put(85,-2.5){\line(0,1){5}}
\put(85,-2.5){\line(1,0){40}}
\put(85,2.5){\line(1,0){40}}
\put(125,-2.5){\line(0,1){5}}
\put(125,0){\vector(1,0){35}}
\put(160,0){\vector(1,0){5}}
\put(165,0){\line(1,0){5}}
\put(35,0){$a$}
\put(180,0){$c$}
\put(84,1){\line(-5,2){10}}
\put(72,7.5){\line(-5,2){10}}
\put(60,15){\line(-5,2){10}}
\put(125,-10){\line(0,1){7}}
\put(125,-20){\line(0,1){7}}
\put(125,-30){\line(0,1){7}}
\put(125,-40){\line(0,1){7}}
\put(45,-40){\vector(1,0){15}}
\put(60,-40){\line(1,0){65}}
\put(125,-42.5){\line(0,1){5}}
\put(125,-37.5){\line(1,0){40}}
\put(125,-42.5){\line(1,0){40}}
\put(165,-42.5){\line(0,1){5}}
\put(165,-42.5){${\bf >}$}
\put(35,-40){$b$}
\put(185,-40){${\bf p}$}
\put(110,-65){\makebox(0,0){\bf b}}
\put(10,-95){\makebox(0,0){\bf Fig. 1}}
\end{picture}\\
The solid lines in Fig.1 
describe the behaviour of two electrons. First both
of them are in the ground state of the atom. The  lines
marked $a$ and $b$ describe the corresponding atomic states. Then the electron
$a$ is ionised. The beginning of the ionization process 
is shown by the sloped
dashed line representing the absorbed photon. 
The propagation of the ionized electron in the laser field, when it 
absorbs another photons, is shown by the 
double solid lines.
Collision of the ionized electron with the parent atomic particle
is shown by the  vertical dashed line representing the Coulomb 
interaction between the electrons. The state marked
$c$ describes the ion excitation which can belong  either to the discrete 
or to the continuum spectrum. In both cases it 
can strongly interact with the laser  field. 
The diagram b describes the exchange process.

The central question of the problem is whether this mechanism 
can work at all. 
It arises because in the process of ATI the ionized
electron could go far outside the atomic particle and therefore
inelastic collision with it could seem improbable. 
To prove that this mechanism works at all
we have  to fulfil  an accurate calculation.
It seems very complicated because we are to consider  the perturbation
theory over the electron interaction when the 
behaviour of the electrons is strongly influenced by the 
time-dependent laser field.
Fortunately, the adiabatic nature of the problem results in a very
important property of the two-electron amplitude: it may be presented
as a product of the amplitude of single-electron ionization and the
amplitude of electron ion impact in the presence of a laser field.
Both latter amplitudes can be reliably calculated. As a result there appears
the possibility of reliable $ab~inicio$ calculation of the two-electron
process. 

Factorization of the two-electron amplitude is one of the most
important results of the paper. The calculations proving it 
are fully presented  because problems of the type considered 
are not well-known in literature. 
We will be forced also to consider briefly two 
related problems: single-electron
ionization and the scattering problem.

Factorization of the amplitude makes it possible to find a reliable
estimation for the probability. It permits us to prove 
that the considered mechanism of two-electron ionization
works well. There is a high probability for the  first ionized electron 
to  return to the parent atomic particle.
 The first ionized electron can  be considered
as a kind of antenna, it absorbs energy from the external field and
transfers it to the ion. This physical idea was first considered
in (Kuchiev 1987) but the corresponding calculations
were not published.

The electric field is assumed to be linearly polarized
\begin{equation} \label{F}
{\bf F}(t)={\bf F} \cos \omega t~.
\end{equation}
The frequency $\omega$ of the field is supposed to be low, $\omega<<I$,
where $I$ is the ionization potential.
The Keldysh adiabaticity parameter
$\gamma$ 
\begin{equation} \label{gamma}
\gamma=\frac{\omega \sqrt{2I} }{F}~
\end{equation}
is considered as arbitrary.
The first ionised electron, which plays so crucial role in the
considered picture, will be described neglecting the static Coulomb
field created by the parent atomic particle. 
A similar description was used in the Keldysh theory
of single-charged ion formation 
(Keldysh 1965, Reiss 1980,1987) .
One can consider it a reasonably good approximation when
the non-resonant processes are important, for example
for the two-electron ionization of noble gases.
This approximation should be even more reliable 
for the problem of two-electron detachment from negative ions.

In Section II the general formulae describing  the single-electron ionization 
problem as well as the two-electron problem are presented. Section III 
contains the prove of factorization of the  two-electron amplitude.
Section IV presents the necessary consideration of the  scattering problem.
In Section V a qualitative physical discussion of the physical ideas 
governing the process of two-electron ionization  
is given. Section VI presents the estimation of the probability of the 
two-electron process for  different regions of adiabatic parameter.
ATI is shown to play the major role when $\gamma>1$.
Section VII gives an illustrative example: the double ionization of He atom
by a 780 nm laser field is briefly considered.

\section{ multi-photon processes}

In this section  a formalism to describe
the multi-photon processes 
when an atom is placed in a 
laser field is developed. The  general
expression for the amplitude $A_{fi}$ 
of an adiabatic process in the field 
is
\begin{equation}\label{afi}
A_{fi} = \frac{1}{T}\int_{0}^{T}
\langle \Psi_f (t) | V_F(t) | \psi_i (t)\rangle ~.
\end{equation}
The wave function $| \psi_i (t)\rangle $ describes the
initial state of the atom taking no account
of  the interaction of the atom with the laser
field. This interaction is described by the operator $V_F(t)$
\begin{equation}\label{vt}
V_F(t) = -{\bf r F}\cos{\omega t}~,
\end{equation}
where $ {\bf r}=\sum {\bf r}_i $ is the dipole moment. We 
use the $r$-form for interaction with the laser field.
The wave function $\langle \Psi_f (t) |$ describes the  final
state of  the reaction taking full account of the interaction
of the atom with the laser field. It makes
the amplitude (\ref{afi})  describe the multi-photon, 
nonlinear process.
Eq.(\ref{afi}) looks 
similar to the usual expression for stationary processes
except for the fact that in (\ref{afi}) there is 
the nontrivial integration over the
period of time $T= 2\pi/\omega$, where $\omega$ is the
frequency of the laser field. 

\subsection{Single-electron ionization}
First let us consider shortly the well-known Keldysh
problem of multi-photon 
single-electron ionization (Keldysh 1965).
focusing our attention on
those points which  will be used in the following 
consideration of many-electron processes.
The initial
wave function for the single-electron process is taken as
\begin{equation}\label{phia}
\psi_i (t)=\Phi_{a}({\bf r},t)=
\phi_a ({\bf r})\exp{(-i E_a t)}~,
\end{equation}
where $\phi_a({\bf r})$ is the wave function of  the
ground state of the atom, and $E_a$ is the atomic binding 
energy.
The wave function of the final state must include the 
interaction of the photo-electron with the external field. If one
neglects the static ion field in the final state 
then this wave function is 
equal to the Volkov wave function (Volkov 1935)
$\Psi_f (t)=\Phi_{\bf p}({\bf r},t)$,
\begin{eqnarray}\label{phip}
&&\Phi_{\bf p}({\bf r},t)=
\exp{\left\{ i \left [  ( {\bf p}+{\bf k}_t){\bf r}-
\int_{0}^{t} E_{{\bf p}}(\tau) 
d \tau  +\frac{{\bf p F}}{\omega^2} \right ] \right \} };\\ \label{ktEt}
&&{\bf k}_t = \frac{ {\bf F} }{\omega} \sin{\omega t},~~~~
E_{{\bf p}}(\tau) =\frac{1}{2} ({\bf p}+{\bf k}_\tau )^2~, 
\end{eqnarray}
which satisfies the Schrodinger equation
\begin{equation}\label{shrH0}
i\frac{\partial}{\partial t}\Phi_{\bf p}({\bf r},t)=
\left ( -\frac{1}{2}\Delta - 
{\bf r F}\cos{\omega t}\right ) \Phi_{\bf p}({\bf r},t)~.
\end{equation}
The additional phase ${\bf p F}/\omega^2$ in Eq.(\ref{phip})
is chosen to provide a convenient property 
\begin{eqnarray}\label{PhiP}
&&\Phi_{\bf p}({\bf r},t+T/2)=\Phi_{-{\bf p}}(-{\bf r},t)
\exp \{-i \bar E_{\bf p} T/2 \},\\ \nonumber
&&\bar E_{\bf p}=\frac{1}{T}\int_{0}^{T}
E_{\bf p}(t)dt=
\frac{ p^2}{2}+\frac{ F^2}{4 \omega^2}~,
\end{eqnarray}
which is essential for consideration of  the parity conservation law,
 see Eq.(\ref{par}) below.
Substituting
(\ref{phia}),(\ref{phip}) in (\ref{afi})
one gets the known expression for the amplitude 
$A^{(e)}({l,\bf p})$
of single-electron ionization when $l$ quanta are absorbed and
the final-state momentum of the electron is ${\bf p}$:
\begin{equation}\label{kel}
A^{(e)}(l;{\bf p}) = 
 \frac{1}{T} \int_{0}^{T}
\langle \Phi_{{\bf p}} (t)| V_F(t) | \Phi_a (t)\rangle=
\frac{1}{T}\int_{0}^{T}
\langle {\bf p}+{\bf k}_t | V_F(t) | \phi_a \rangle
\exp \left \{\frac{i}{\omega} S(\omega t) \right \} ~.
\end{equation}
Here the usual definition 
$|{\bf p}\rangle  = \exp \{i {\bf p r} \}$ is used and $S(x)$ is
\begin{equation} \label{Sx}
S(x)= \int_{0}^{x} \frac {1}{2} \Bigl({\bf p}+
\frac{{\bf F}}{\omega} \sin x \Bigr)^2dx-E_{a} x~-\frac{{\bf p F}}{\omega}.
\end{equation}
The energy conservation law for the process of ionization reads
\begin{equation}\label{enkel}
E_a+l\omega= \bar E_{\bf p}~.
\end{equation}
The parity conservation law for the ionization process
manifests itself as a condition for the amplitude
\begin{equation}\label{par}
A^{(e)}(l;-{\bf p})=(-1)^l P_a A^{(e)}(l;{\bf p})~,
\end{equation}
where $P_a=\pm1$ is the parity of the atomic state $a$.
It is easy to verify using Eq.(\ref{PhiP}) 
that the amplitude Eq.(\ref{kel}) satisfies this condition.

The integrand in Eq.(\ref{kel}) contains a large phase $\sim 1/\omega$
  and therefore the
steepest descent method is applicable. The saddle points, which must be taken 
in the upper semiplane of the complex plane  $x=\omega t,~
{\rm Im}~x>0,$ in the region $0\le {\rm Re}~x \le 2 \pi$ satisfy the equation
\begin{equation} \label{S'}
S'(x)=\frac{1}{2} \Bigl({\bf p}+ \frac{{\bf F}}{\omega}
\sin x \Bigr)^2-E_{a}=0~,
\end{equation}
which results in the following condition
\begin{equation} \label{uvpkap}
\sin x =\frac{\omega}{F}
\left [ - p_{||} \pm i ( \kappa^2+p_{\perp}^2 )^{1/2}\right ]~,
\end{equation}
where $p_{||}$ and $p_{\perp}$ are the longitudinal and transverse
components of the vector ${\bf p}$ in respect to the field
${\bf F}$ and $\kappa =\sqrt{2|E_a|}$.  
It follows from
Eq.(\ref{S'}) that there are two  saddle points $x_1,x_2$
in the region of interest
which 
we will label in such a way that
\begin{equation}\label{cosg0}
{\rm Re}\cos x_1 >0, ~{\rm Re}\cos x_2 < 0;~~ 
{\rm Im}~x_1,{\rm Im}~x_2\ge0,~~0\le {\rm Re}~x_1,{\rm Re}~x_2\le2 \pi~.
\end{equation}
Calculating
the integral in Eq.(\ref{kel}) by the saddle-point
method one finds
\begin{eqnarray} \label{atij}
A^{(e)}({l,\bf p})&=&\sum_{\sigma=1,2}
A^{(e)}_\sigma ({l,\bf p}),\\ \label{ati}
A^{(e)}_\sigma ({l;\bf p})&=&
 \frac{1}{T} \int \limits_{C_\sigma}
\langle \Phi_{{\bf p}} (t)| V_F(t) | \Phi_a (t)\rangle dt=\\ \nonumber
&&\sqrt {\frac{i~ 
\omega}{2 \pi S''(x_\sigma )}} \exp \left\{ \frac{i}{\omega}
S(x_\sigma ) 
\right\}
\Big \{ \langle {\bf p}+{\bf k}_t| V_F(t) | \phi_a \rangle \Big \}
_{(t=x_\sigma /\omega )}~,
\end{eqnarray}
where $A^{(e)}_\sigma ({l,\bf p}), \sigma =1,2,$
is the contribution to the amplitude
of the  saddle point $x=x_\sigma$ satisfying Eqs.(\ref{S'}), 
(\ref{cosg0}) and $C_\sigma$ is the part of the integration 
contour in the $t$ plane which crosses the saddle point $x_\sigma$.
(Note that evaluation of 
$\langle {\bf p}+{\bf k}_t| V_F(t) | \phi_a \rangle$ 
needs  accuracy 
because of the singular nature of this matrix element
for complex $t$.)
It is easy to verify that the amplitudes $A^{(e)}_\sigma({l;\bf p})$
satisfy  the condition
\begin{equation}\label{AeP}
A^{(e)}_2 (l;-{\bf p})=(-1)^l P_a A^{(e)}_1({l;\bf p})~,
\end{equation}
which shows explicitly that the amplitude
Eq.(\ref{atij}) obeys the parity conservation law Eq.(\ref{par}).
The absolute values of 
$A^{(e)}_1({l;\bf p})$ and $A^{(e)}_2({l;\bf p})$
are equal. In contrast, 
their phases differ  substantially and this
phase difference
depends on the field $F$. In order to check it out consider
small $ {\bf p}$, $p^2 \ll \kappa^2$. For this case the phase 
difference is easily found to be
$2 \sqrt{1+\gamma^2}Fp_{||}/\omega^2$. The estimation
for the smallest $p_{||}$ is $p_{||}\sim \sqrt{\omega}$. Therefore
the  estimation for the phase difference 
is $\sim 2 (F^2/\omega^2+\kappa^2)^{1/2}/\sqrt{\omega}\gg 1$. 
The large phase difference makes  interference between
$A^{(e)}_1({l;\bf p})$ and $A^{(e)}_2({l;\bf p})$
not important
for   many cases. Then calculating the probability one
can suppose that
\begin{equation}\label{prob2A}
|A^{(e)}({l;\bf p})|^2 = 2|A^{(e)}_1({l;\bf p})|^2~.
\end{equation}

\subsection{Many-electron processes}
Let us develop an approach to the two-electron processes 
in a way similar to the single-electron  ionization.
The main purpose of this section is to obtain the analytical expression
describing the Feynman diagrams in Fig.1.
The initial two-electron wave function
may be chosen as a product of the single-electron
wave functions
\begin{equation}\label{psi2i}
\psi_i({\bf r}_1,{\bf r}_2,t)=
\phi_a ({\bf r}_1)\phi_b ({\bf r}_2) 
\exp{\lbrace -i (E_a+E_b) t)\rbrace}~.
\end{equation}
This form of the wave function  
neglects the initial-state correlations, but they do not play
a role. The most interesting for our purpose is the 
final-state interaction. To simplify the 
presentation  we do not take into account
explicitly the symmetry of the wave function with respect
to permutation of electrons though it can be restored
as discussed below. At the moment let us consider
the two electrons as if they are distinguishable. The
electron whose coordinates are ${\bf r}_1$ will be considered
as an atomic electron. It will also be referred to as ``the 
first'' electron. The electron with 
coordinates ${\bf r}_2$ will represent the degrees of freedom
of the single-charged ion. It will be called ``the second''
electron.
The wave function $\phi_a ({\bf r}_1)$ describes
the ground state of the atom, $E_a$ is the binding
energy.  The function $\phi_b ({\bf r}_2)$ is to  be 
considered as the wave function of the ground state of the
single-charged ion, $E_b $ being the binding
energy of this state.  Note that the removal of the first
electron is an adiabatic process, see Section A. Hence
the ion remains mainly in the ground state after
single-electron ionization.
We are interested in the interaction of the atom with
the laser field. Therefore we consider the interaction
of the first electron with the field. In particular
we take into account  the excitation of this electron into 
continuum states. In contrast, the direct interaction
of the ion in the ground state with the laser field
is strongly suppressed due to a high ionization potential of 
the ion. That is why we will  neglect direct interaction
of the second electron with the laser field
when  it occupies the ground state. 
Certainly  this interaction will be considered
for the excited states.
Thus the only possibility to be considered
for  the excitation of the second electron
comes from its interaction with the first one. 

If we neglect the two-electron interaction 
then the final state wave function can be chosen as  
a product of the Volkov wave function for  the
photoelectron and the ion ground state wave function
\begin{equation}\label{psi2f0}
\Psi_{fb}({\bf r}_1,{\bf r}_2,t)=
\Phi_{\bf p}({\bf r}_1,t)
\phi_{b}({\bf r}_2)\exp{ \lbrace -i E_b t \rbrace }.
\end{equation}
Substituting Eqs.(\ref{psi2i}),(\ref{psi2f0}) into 
Eq.(\ref{afi}) we reproduce Eq.(\ref{kel}) for       
the single-electron process.

Consider now the excitations of the single-charged ion
 into discrete or continuum states. 
The later case
describes the doubly charged ion formation.
The simplest wave function describing the excitation
is given by the product of the wave function of the 
ionised electron and the wave function of the excited ion
\begin{equation}\label{psi2f0c}
\Psi_{fc}({\bf r}_1,{\bf r}_2,t)=
\Phi_{\bf p}({\bf r}_1,t)
\Phi_{c}({\bf r}_2,t)~.
\end{equation}
Here we describe the wave function of the first electron
in the continuum by the Volkov wave function (\ref{phip}).
The function $\Phi_{c}({\bf r}_2,t)$ describes the excited
state $c$ of the ion interacting with the laser field. It 
satisfies the time-dependent Schrodinger equation
\begin{equation}\label{Hphic}
i\frac{\partial}{\partial t}\Phi_c({\bf r}_2,t)=
\left ( H_{ion}({\bf r}_2)
- {\bf r}_2{\bf F}\cos{\omega t} \right )
\Phi_c({\bf r}_2,t)~,
\end{equation}
where $H_{ion}$ is the Hamiltonian  of the single-charged ion.
We will consider the discrete as well as the continuum state
excitations. The wave function of the  discrete state, if it is 
well separated from the other states,
may be found in an adiabatic approximation
\begin{equation}\label{phin}
\Phi_c({\bf r}_2,t)=\phi_c({\bf r}_2,t)\exp{\left \{
-i \int_{0}^{t}E_c(\tau)d\tau \right \}}~.
\end{equation}
Here $\phi_c({\bf r}_2,t)$ is the wave function
satisfying the stationary Schrodinger equation
\begin{equation}\label{phins}
E_c(t)\phi_c({\bf r}_2,t)= (H_{ion}({\bf r}_2)-
{\bf r}_2{\bf F}\cos{\omega t}) \phi_c ({\bf r}_2,t)~,
\end{equation}
with the time-dependent ``energy'' $E_c(t)$. When the second
order of perturbation theory over the field is applicable then
\begin{equation}\label{Ent}
E_c(t)=E_c - \frac{1}{2} \alpha_c(\omega){ F}^2 
\cos^2\omega t~,
\end{equation}
where $E_c$ is the energy of the stationary ion state $c$
and $\alpha_c(\omega)$ is the dynamical polarizability
of this ion state. The excitation of the ion into the continuum
will be described with the help of the Volkov wave
functions (\ref{phip}). It is convenient to consider
the discrete state
excitations and the excitations into the continuum on 
equal ground. 
Note that if we neglect the ion static field in 
Eq.(\ref{phins}) then this equation is reduced
to Eq.(\ref{shrH0}) describing  the continuum states
of the ion.
Therefore we can consider the discrete states of the ion
and its continuum states 
using the same set of Eqs.(\ref{phin}),(\ref{phins}) 
for the wave function
and keeping in mind that $H_{ion}({\bf r}_2)\approx
-(1/2)\Delta_2$
if a state in the continuum  is considered. We will suppose that 
the phase in the definition of $\Phi_c({\bf r},t)$ is chosen in such
a way that
\begin{eqnarray}\label{cP}
&&\Phi_c({\bf r},t+T/2)=\Phi_{P(c)}(-{\bf r},t)\exp \{-i \bar E_c T/2\},
\\ \nonumber
&&\bar E_c= \frac{1}{T}\int_0^T E_c(t)dt~,
\end{eqnarray}
where $\Phi_{P(c)}$ is the wave function of the 
state with the quantum numbers  obtained by applying
the inversion operator $P$ to the quantum numbers of the state $c$.
 The definition Eq.(\ref{cP}) agrees with
Eq.(\ref{PhiP}). The possibility to satisfy Eq.(\ref{cP}) 
follows from the Schrodinger Eq.(\ref{Hphic}).

The excitation of the ion takes place due to
a mixing of  the 
wave function (\ref{psi2f0c}) describing the excitation
of the ion 
with  the wave function  (\ref{psi2f0}) describing its
ground state. In order to find this mixing 
consider the time-dependent Schrodinger
equation for the final state wave function
\begin{equation}\label{shr2}
-i\frac{\partial}{\partial t}\Psi_{f}^{*}
({\bf r}_1,{\bf r}_2,t)=
\left ( H_0({\bf r}_1,{\bf r}_2,t)+\frac{1}{r_{12}} \right )
\Psi_{f}^{*}({\bf r}_1,{\bf r}_2,t)~.
\end{equation}
Here
\begin{equation}\label{H02}
H_0({\bf r}_1,{\bf r}_2,t)=-\frac{1}{2}\Delta_1
- {\bf r}_1{\bf F}\cos{\omega t}+H_{ion}({\bf r}_2)
- {\bf r}_2{\bf F}\cos{\omega t}~
\end{equation}
is the Hamiltonian describing the considered
two-electron system
when the interaction between the electrons is neglected.
This interaction 
is accounted for in Eq.(\ref{shr2})
by the term $1/r_{12}$.
Consider this interaction in the first order of perturbation
theory. We are looking for the final state
wave function which
in the future, $t \rightarrow \infty$, 
gives the excited ion described by the 
wave function Eq.(\ref{psi2f0c}). Therefore the later
wave function gives the zero approximation
\begin{equation}\label{zerpsi}
\Psi_{f}({\bf r}_1,{\bf r}_2,t)\approx
\Psi_{f}^{(0)}({\bf r}_1,{\bf r}_2,t)=  
\Phi_{\bf p}({\bf r}_1,t)
\Phi_{c}({\bf r}_2,t)~.
\end{equation}
Then the 
correction $\delta \Psi_{f}^{*}({\bf r}_1,{\bf r}_2,t)$
to  the wave function caused by the
interaction between the electrons
satisfies the equation
\begin{equation}\label{delpsi}
\left( -i\frac{\partial}{\partial t} -
 H_0({\bf r}_1,{\bf r}_2,t)\right )
\delta \Psi_{f}^{*}({\bf r}_1,{\bf r}_2,t)=\frac{1}{r_{12}}
\Psi_{f}^{(0)*}({\bf r}_1,{\bf r}_2,t)~.
\end{equation}
We are to find $\langle \delta \Psi_{f}|$ from Eq.(\ref
{delpsi}) and use it in Eq.(\ref{afi}) for the amplitude.
This procedure is
simplified due to the fact that we
neglect the interaction of the ion in the ground state
with the laser field. This means that 
we neglect the transitions from the state
described by the wave function $\phi_b({\bf r}_2)$.
It makes the operator $V_F(t)$ in Eq.(\ref{afi}) 
 independent on the coordinate ${\bf r}_2$,
$V_F(t)=-{\bf r}_1 {\bf F} \cos \omega t$.
Thus the integration over ${\bf r}_2$ in (\ref{afi})
is reduced to the projection of the final state
wave function on $\phi_b({\bf r}_2)$ which is convenient to evaluate
in advance. 
Let us denote it as $\Psi_{f}^{*}({\bf r}_1,t)$
\begin{equation}\label{psifn}
\Psi_{f}^{*}({\bf r}_1,t)=\int 
\delta \Psi_{f}^{*}({\bf r}_1,{\bf r}_2,t)\phi_b({\bf r}_2)
\exp{ \{-i E_b t \}} d^3 r_2~.
\end{equation}
From Eq.(\ref{delpsi}) we find that it satisfies the equation
\begin{equation}\label{shr1in}
\left ( -i\frac{\partial}{\partial t}
+\frac{1}{2}\Delta_1 + 
{\bf r}_1{\bf F}\cos{\omega t} \right )
\Psi_{f}^{*}({\bf r}_1,t)=
V_{cb}({\bf r}_1,t)\Phi_{\bf p}^{*}({\bf r}_1,t)~,
\end{equation}
where $V_{cb}({\bf r}_1,t)$ is 
\begin{equation}\label{vnb}
V_{cb}({\bf r}_1,t)=\exp{\left \{ i \left ( 
\int_{0}^{t}E_c (\tau) d \tau-E_b t \right )\right \} }
\int \phi_c^*({\bf r}_2,t)
\frac{1}{r_{12}}
\phi_b ({\bf r}_2)d^3 r_2~.
\end{equation} 
The solution of Eq.(\ref{shr1in}) reads
\begin{equation}\label{solpsifn}
\Psi_{f}^{*}({\bf r}_1,t)=\int_{t}^{\infty}d t^{'} \int
\Phi_{\bf p}^{*}({\bf r}^{'}_1,t^{'})
V_{cb}({\bf r}^{'}_1,t^{'}) G({\bf r}^{'}_1,t^{'};{\bf r}_1,t)
d^3r^{'}_1~.
\end{equation}
Here we introduce the time-dependent retarding  
Green function $G({\bf r}^{'}_1,t^{'};{\bf r}_1,t)$ describing 
the electron propagation in the laser field. It satisfies the equations
\begin{eqnarray}\label{shgr}
\left ( -i\frac{\partial}{\partial t}
+\frac{1}{2}\Delta + 
{\bf r F}\cos{\omega t} \right )
& &G({\bf r}^{'},t^{'};{\bf r},t)= \delta ({\bf r}^{'}-
{\bf r})\delta (t^{'}-t), \\ \nonumber
& &G({\bf r}^{'},t^{'};{\bf r},t)=0,~~~ 
{\mbox {if}}~~ t^{'}<t~.
\end{eqnarray}
Now from Eqs.(\ref{solpsifn}),(\ref{afi})
we find the amplitude $A^{(2e)}(n; {\bf p},c)$
describing the two-electron process
when $n$ quanta are absorbed,
one electron occupies the final state ${\bf p}$
in the continuum and the other 
  occupies the discrete or continuum
state $c$ 
\begin{equation}\label{Ani}
A^{(2e)}(n; {\bf p},c)=\frac{1}{T}\int \limits_{0}^{T}dt
\int \limits_{t}^{\infty}dt^{'}
\langle \Phi_{\bf p}(t{'})|V_{cb}(t^{'})G(t^{'},t)V_F(t)|
\Phi_a(t)\rangle ~.
\end{equation}
Integration over the
variables ${\bf r}_1,{\bf r}_1^{'}$ is presented in (\ref{Ani})
in the symbolic form, integration over the variable
${\bf r}_2$ is performed evaluating $V_{cb}$,
see Eq.(\ref{vnb}). Note that one can restore
the symmetry with respect to permutation of the
electrons if along with the matrix element
$V_{cb}$ the exchanged matrix element is considered as
shown in Fig.1 b.

The energy conservation law for the process
considered reads
\begin{equation}\label{E2con}
E_a+ E_b+n\omega= 
\bar E_{\bf p}+ \bar E_c~,
\end{equation}
where $n$ is the total number of absorbed quanta. 
Eq.(\ref{E2con}) is similar to Eq.(\ref{enkel}) for 
the single-electron process and 
can be justified in a similar manner, see Section III. 

Eq.(\ref{Ani}) describes the ionization of the atom which is accompanied
by the excitation of the ion into the discrete or continuum
state. It gives the analytical expression for the diagram
a in Fig.1. Let us formulate the main
 assumptions made in its derivation.
\\ 1.The transitions from the ion ground state caused by the
laser field are neglected.
\\ 2.The wave functions in the continuum are described
by the Volkov wave functions.
\\3.Interaction between the electrons is considered
in the first order of perturbation theory.

\section{factorization of the amplitude} 
Direct calculation of the amplitude (\ref{Ani}) is
complicated due to the adiabatic nature of the 
process under consideration which makes the integrand exhibit a 
strong variation with respect to the time variables
$t,t^{'}$. Fortunately the same adiabatic nature
provides a very useful   property of the amplitude: 
the amplitude may be considered as a product
of the single electron photo-ionization amplitude and the
amplitude of electron-ion inelastic scattering. We will call
this property  factorization. For the first time this idea
was considered in (Kuchiev 1987).

In order to prove factorization
let us first rewrite
Eq.(\ref{Ani}) in the more convenient form given below by 
Eqs.(\ref{Afin}),(\ref{Aden}),(\ref{Arr}). With this purpose
consider  representation for  the Green function
through the Volkov wave functions
\begin{equation}\label{grfu}
G({\bf r}^{'},t^{'};{\bf r},t)=
-i \theta( t^{'}-t)
\int \Phi_{\bf q}({\bf r}^{'},t^{'})
\Phi_{\bf q}^*({\bf r},t)\frac{d^3 q}{(2\pi)^3}~.
\end{equation}
This equality can easily be justified with
the help of Eqs.(\ref{shgr}),(\ref{phip}). Using
it one can rewrite Eq.(\ref{Ani}) 
\begin{equation}\label{Aspec}
A^{(2e)}(n, {\bf p},c)=(-i)\frac{1}{T}\int \limits_{0}^{T}dt 
\int \limits_{t}^{\infty}dt^{'}
\int \frac{d^3 q}{(2\pi)^3}
\langle \Phi_{\bf p}(t{'})|V_{cb}(t^{'})|\Phi_{\bf q}(t^{'})
\rangle 
\langle \Phi_{\bf q}(t)|V_F(t)|\Phi_a(t)\rangle~.  
\end{equation}
It is important to distinguish in Eq.(\ref{Aspec})
the periodic functions
of $t^{'}$ from the factors having the
form $\exp \{ iEt^{'} \}$ which give
the monotonic contribution to the phase.
The function $\phi_c({\bf r}_2,t{'})$ defined in Eq.(\ref{phins})
is periodic. The last factor in the right-hand side
of (\ref{phin}) can be presented as
\begin{equation}\label{evEc}
\exp{\left \{ i  
\int_{0}^{t}E_c (\tau) d \tau \right \} }=
\exp{\lbrace i \bar E_c t\rbrace }
\exp{\left \{ i  
\int_{0}^{t}(E_c (\tau)-\bar E_c  )d \tau \right \} }~.
\end{equation}
Then the matrix element $V_{cb}({\bf r}_1,t)$, see Eq.(\ref{vnb}), 
can be presented as
\begin{equation}\label{VbarE}
V_{cb}({\bf r},t^{'})=
\exp{\lbrace i (\bar E_c-E_b) t^{'} \rbrace }
U_{cb}({\bf r},t^{'})~,
\end{equation}
where
\begin{equation}\label{Ucb}
U_{cb}({\bf r}_1,t^{'})=
\exp{\left \{ i \left (
\int_{0}^{t^{'}}(E_c (\tau ) -\bar E_c) d \tau \right ) 
\right \} }
\int \phi_c^*({\bf r}_2,t^{'})
\frac{1}{r_{12}}
\phi_b ({\bf r}_2)d^3 r_2~
\end{equation}
is a periodic function of $t^{'}$.
Similar representation for the Volkov functions 
(\ref{phip}) reads
\begin{eqnarray}\label{phibarE}
\Phi_{\bf p}({\bf r},t)&=&
\chi_{\bf p}({\bf r},t)\exp{ \lbrace -i \bar E_{\bf p}t
\rbrace} ~,\\ \label{chi}
\chi_{\bf p}({\bf r},t)&=&
\exp{\left \{ i \left [ ( {\bf p}+{\bf k}_t){\bf r}-
\int_{0}^{t} \left ( E_{{\bf p}}(\tau)-
\bar E_{\bf p} \right ) 
d \tau + \frac{ {\bf p F} }{\omega^2} \right ] \right \} }~.
\end{eqnarray}
From (\ref{Aspec}),(\ref{VbarE}),(\ref{phibarE}),(\ref{chi})  we find
\begin{eqnarray}\label{Anirrq}
A^{(2e)}(n; {\bf p},c)=(-i)\frac{1}{T}
\int \limits_{0}^{T}dt\int \limits_{t}^{\infty}dt^{'}
\int \frac{d^3 q}{2\pi^3}
\langle \chi_{\bf p}(t{'})|U_{cb}(t^{'})|\chi_{\bf q}(t{'})
\rangle 
\langle \chi_{\bf q}(t)|V_F(t)|\phi_a\rangle  \times 
\\ \nonumber
\exp{ \lbrace i\lbrack(\bar E_{\bf p}+\bar E_c-E_b-
\bar E_{\bf q})t^{'}+
(\bar E_{\bf q}-E_a)t \rbrack \rbrace}~.
\end{eqnarray}
Now it is useful to take into account the identity
\begin{equation}\label{idenfe}
\int_{t}^{\infty}d t^{'}\exp{ \lbrace i Et^{'} \rbrace } 
f(t^{'})=
i \sum_{m=-\infty}^{\infty} \frac{1}{T}\int_{0}^{T} d t^{'}
f(t^{'}) \exp{ \lbrace i\lbrack (E-m\omega)t+
m\omega t^{'}\rbrack \rbrace}\frac{1}{E-m\omega+i 0}~,
\end{equation}
which is valid for any periodic function $f(t)=f(t+T)$. 
The identity can easily be checked out with the help of 
Fourier expansion  $f(t)=\sum_{m} f_m 
\exp{\lbrace -im\omega t \rbrace }$.
Using Eq.(\ref{idenfe})  we find from Eq.(\ref{Anirrq})
\begin{equation}\label{Afin}
A^{(2e)}(n; {\bf p},c) = \sum_{l}A^{(2e)}(n,l; {\bf p},c)~,
\end{equation}
where 
\begin{eqnarray}\label{Aden}
A^{(2e)}(n,l; {\bf p},c) = \frac{1}{T^2}\int_{0}^{T}dt dt^{'}
&&\int \frac{d^3 q}{2\pi^3}
\frac{\langle \chi_{\bf p}(t{'})|U_{cb}(t^{'})|\chi_{\bf q}
(t{'})\rangle 
\langle \chi_{\bf q}(t)|V_F(t)|\phi_a\rangle }
{\bar E_{\bf p}+\bar E_c-E_b-
\bar E_{\bf q}-(n-l)\omega+i0}\times\\ \nonumber
&&\exp{ \lbrace i\lbrack (\bar E_{\bf p}+\bar E_c-E_b-
 E_a -(n-l)\omega) t+(n-l)\omega t^{'}\rbrack \rbrace} ~.
\end{eqnarray}
This formula looks similar to the usual representation for
amplitudes of stationary processes.
It is clear that the factor 
$\langle\chi_{\bf q}(t)|V_F(t)|\phi_a\rangle$ in 
Eq.(\ref{Aden}) describes 
the matrix element of the transition of atomic electron into 
the continuum, while 
$\langle \chi_{\bf p}(t{'})|U_{cb}(t^{'})|\chi_{\bf q}
(t{'})\rangle $ is the matrix element of
 the inelastic scattering of the photoelectron on the
ion. The non-stationary nature of the process
manifests itself - there are  nontrivial
integrations over the time variables in Eq.(\ref{Aden}). 
The integration over $t$
results in particular
in the energy conservation law
$E_a+ E_b+n\omega= 
\bar E_{{\bf p}}+
\bar E_c$,  Eq.(\ref{E2con}).
To simplify notation in the following consideration the index of 
summation $l$ in Eqs.(\ref{Afin}),(\ref{Aden}) is chosen 
differently from Eq.(\ref{idenfe}), $l=n-m$.
The physical interpretation of the number $l$
comes from  integration over $t$ which gives 
 the $l$-th harmonic of the matrix element
of single-electron ionization. Therefore $l$ is to
be considered as the   number of quanta absorbed 
during the single-electron ionization.
Similarly, the integration over $t^{'}$ shows that $n-l$ is
the number of quanta absorbed in the 
process of inelastic collision of the photoelectron with
the ion. At this stage of consideration this
interpretation is not of great importance, but later,
when we prove factorization, it will become essential.

Our next step is to simplify Eq.(\ref{Aden}). With this purpose
note the following.
First, integration over $t$ describes an adiabatic
transition and therefore can be evaluated by the steepest descent method
at the saddle points situated in the complex plane $t$.
Second,
the wave function $\chi_{{\bf q}}({\bf r},t)$, see  Eq.(\ref{chi}),
depends on ${\bf q}$ in a very simple way. It is an exponent of a 
linear form of ${\bf q},~
\chi_{{\bf q}}({\bf r},t) = \exp \{ i[ {\bf q R}+ \alpha ]\}$,
${\bf R}$ and $\alpha$ being ${\bf q}$-independent. This allows us
to reduce integration over $d^3q$  to the residue at the pole
arising due to a node of the denominator 
$\bar E_{\bf q}=\bar E_{\bf p}+\bar E_c-E_b-(n-l)\omega$.
These facts make it convenient to deform the contour of integration over $t$
into the upper semiplane of the complex plane $t$
to bring it close to  the saddle points
\begin{equation}\label{defo}
\int_{0}^{T} \exp \{il\omega t\}
\langle \chi_{\bf q}(t)|V_F(t)|\phi_a \rangle dt=
\int \limits_{C_2+C_1+C_0} \exp \{il\omega t\}
\langle \chi_{\bf q}(t)|V_F(t)|\phi_a \rangle dt~.
\end{equation}
The integrand here is similar to one in the single-electron
ionization problem, see Section II A. From its  analysis we
know that there are two saddle points of interest $\omega t=x_1,x_2$, see 
Eq.(\ref{cosg0}). We call $C_\sigma,~\sigma=1,2$ 
that part of the contour of integration which is situated
in the vicinity of the point $t=x_\sigma/\omega$. 
The integrand on the 
contour $C_1$ is the function obtained by 
analytical continuation from those regions of the
real axes $t$ where $\cos \omega t >0$. Similarly, the integrand on 
the contour $C_2$ is the analytical function obtained by
 continuation from those regions of the
 real axes where $\cos \omega t <0$.

The contour $C_0$ in 
Eq.(\ref{defo}) describes those parts of the contour of integration 
which are well separated from the saddle points and therefore
give negligible   contribution to the integral, $\int_{C_0}\approx 0$. 
There is  no problem with the end-points $t=0,T$ 
as their contributions compensate for each other
due to the  periodic nature of the integrand. The saddle 
points $\omega t=x_1,x_2$
depend on ${\bf q}$, but it is convenient to choose  $C_1,C_2$ as 
${\bf q}$-independent contours situated  
in the vicinity of the pole
$\bar E_{\bf q}=\bar E_{\bf p}+\bar E_c-E_b-(n-l)\omega$. We can do it
because then for any given $ {\bf q}$ situated 
close to the pole we can deform 
$C_1,C_2$ in such a way that these contours  pass through the saddle points. 
As  mentioned, the vicinity
of the pole is the only important region
 of integration over ${\bf q}$.

Substituting Eq.(\ref{defo}) in Eq.(\ref{Aden}), using the coordinate
representation given in Eq.(\ref{chi}) for the wave functions and integrating
over $d^3q$ we find 
\begin{eqnarray}\label{Arr}
&&A^{(2e)}(n,l; {\bf p},c)=
\frac{1}{T^2}\int \limits_{C_1+C_2}dt\int \limits_{0}^{T}dt^{'}
\int d^3r d^3r^{'}
\chi_{{\bf p}}^{*}({\bf r}^{'},t^{'})
U_{cb}({\bf r}^{'},t^{'})
  (-{\bf rF}\cos{\omega t}) 
\phi_a({\bf r}) \times  
 \\ \nonumber 
&&\left ( -\frac{1 }{2 \pi R} \right )
\exp{ \left \{ i \left [K_{l} R+ l\omega t+(n-l)\omega t^{'} +
{\bf k}_{t^{'}}{\bf r}^{'}-
{\bf k}_{t}{\bf r}-\int_{t}^{t^{'}}
\left ( \frac{1}{2}{\bf k}_{\tau}^2-
\frac{{\bf F}^2}{4\omega^2} \right )
\right ] \right \} }~, 
\end{eqnarray}
where 
\begin{equation}\label{Kmnm}
K_{l} =\left [ 2 \left ( l\omega -
\frac{  F^2 }{4 \omega^2}+ E_a\right ) \right ]^{1/2}  
\end{equation}
is the momentum of the photoelectron after absorption of 
$l$ quanta.
$R = \sqrt{{\bf R}^2}$ in Eq.(\ref{Arr}) is the function of all 
the variables of 
integration 
\begin{equation}\label{R12}
{\bf R}={\bf R}({\bf r},{\bf r}^{'};t,t^{'})=
\int_{t}^{t^{'}} {\bf k}_{\tau} d\tau - {\bf r}^{'} +{\bf r} 
=\frac{{\bf F}}{\omega^2}(\cos \omega t-
\cos \omega t^{'})- {\bf r}^{'} +{\bf r}   ~.
\end{equation}
Eq.(\ref{Arr}) 
is very convenient for evaluating  parameters
governing the process.  First of all note that integration
over the variables ${\bf r},{\bf r}^{'}$ is localised in
the vicinity of the atom. Really,
integration over ${\bf r}$ describes the 
matrix elements responsible for the
ionization process, while integration over ${\bf r}^{'}$
describes the matrix elements of inelastic collision of the
photoelectron with the ion.
These ${\bf r},{\bf r}^{'}$ should be compared with
the amplitude of wiggling of the electron in the laser 
field $F/ \omega^2$. The latter becomes large
even for quite moderate electric fields
\begin{equation}\label{fom}
\frac{F}{\omega^2} \gg 1~.
\end{equation}
Note that this 
inequality may be fulfilled for a big, $\gamma >1$, 
as well as for
small , $\gamma <1$, adiabatic parameter.
Therefore, for the fields satisfying (\ref{fom})
we can assume that 
\begin{equation}\label{abbrr}
\frac{F}{\omega^2} \gg  r,r^{'}~.
\end{equation}
This permits us to simplify $R$ 
\begin{eqnarray}\label{RR0}
& &R\approx R_0 - \frac{ {\bf R}_0\cdot
 ({\bf r^{'}}-{\bf r})}{  R_0 },\\ \label{R0sq}
& &R_0=R_0 (t,t^{'})= \sqrt{{\bf R}_0^2},\\ \label{R0tt}
& &{\bf R}_0 ={\bf R}_0(t,t^{'})=
\frac{{\bf F}}{\omega^2}(\cos \omega t-\cos \omega t^{'})~.  
\end{eqnarray}
As a result the Green function in (\ref{Arr}) becomes simpler
\begin{eqnarray}\label{GR0}
& &\frac{1}{R} 
\exp{\lbrace i K_{l} R \rbrace }\approx
\frac{1}{R_0} 
\exp{ \left \{ i \left (
K_{l} R_0 +  {\bf K}_{l} \cdot
({\bf r}^{'}-{\bf r}) \right ) \right \} },\\ \label{KFF}
& &{\bf K}_{l}=- K_{l} \frac{{\bf R}_0}{R_0} ~.
\end{eqnarray}
Substituting Eq.(\ref{GR0}) in (\ref{Arr}) and using
Eq.(\ref{Kmnm}) to rewrite
 the exponent we find
\begin{eqnarray}\label{facrr}
A^{(2e)}(n,l; {\bf p},c)=
-\frac{1}{2 \pi T^2}\int \limits_{C_1+C_2}dt \int \limits_{0}^{T}dt^{'}
\langle \chi_{{\bf p}}(t^{'})|U_{cb}(t^{'})|{\bf K}_{l}+
{\bf k}_{t} \rangle  \langle {\bf K}_{l}+{\bf k}_{t}|V_F(t)|
\Phi_a(t)\rangle 
\times \\ \nonumber
\frac{1}{ R_0(t,t^{'}) }
\exp \left \{ i \left [K_{l} R_0(t,t^{'})+  \frac{1}{2}K_{l}^2 t+
\left ( \bar E_{{\bf p}}+\bar E_c-E_b  -
\frac{1}{2}K_{l}^2 \right )t^{'}-
\int_{t}^{t^{'}} \frac{1}{2}{\bf k}_{\tau}^2d \tau  
\right ] \right \}~.
\end{eqnarray}
Let us remember that integration over $C_1$ is in
the region where Re$(\cos \omega t) >0$, while integration over
$C_2$ is  in the region where Re$(\cos \omega t) <0$. 
For  inelastic collision the  main contribution
to the integral over $t^{'}$
comes from the vicinities of the points where
\begin{equation}\label{cos0} 
 |\cos \omega t^{'}| \approx 0~,
\end{equation}
see Eq.(\ref{sin0}) in Section IV. (This equality has
a clear physical meaning.
If it is fulfilled then  the wiggling  energy of the 
photoelectron is maximal, $F^2/(2\omega^2)$, providing the 
best opportunity for the ion impact excitation.)
Therefore we can assume 
Re$(\cos \omega t) -\cos \omega t^{'}>0$ when $t\in  C_1$, and 
Re$(\cos \omega t) -\cos \omega t^{'}<0$  when $t\in  C_2$.
The function $R_0(t,t^{'})$ for complex $t$ is to  be considered
as analytical continuation from the real axes. For real $t$
the inequality $R_0(t,t^{'})>0$ is fulfilled. 
 As a result we find
\begin{equation}\label{R0g0}
R_0(t,t^{'})=\frac{F}{\omega^2}\times
\cases {\cos \omega t -\cos \omega t^{'},~~
{\rm if}~t \in C_1~, \cr 
\cos \omega t^{'} -\cos \omega t,~~
{\rm if}~t\in C_2~. \cr }
\end{equation}
Substituting Eq.(\ref{R0g0})  in Eq.(\ref{facrr})
we come to the final expression for the amplitude
\begin{eqnarray}\label{Afactsums}
A^{(2e)}(n,l; {\bf p},c)&=&\frac{ 1}{R_{l}} 
\sum_{\sigma=1,2}A^{(2e)}_\sigma (n,l; 
{\bf p},c), \\ \label{Afactsum}
A^{(2e)}_\sigma(n,l; {\bf p},c)&=&A^{(e)}_\sigma (l; {\bf Q}_{\sigma,l})
A^{(e-2e)}(n-l; {\bf Q}_{\sigma,l};{\bf p},c))~.
\end{eqnarray}
Here $\sigma=1,2$ numerates contributions of the 
two contours $C_1,C_2$ of integration  over $t$.
The quantities $A^{(e)}_\sigma$  are 
\begin{equation}\label{Qsi}
A^{(e)}_\sigma (l; {\bf Q}_{\sigma,l})=\frac{1}{T}
\int \limits_{C_\sigma}\langle \Phi_{{\bf Q}_{\sigma,l}}(t) |V_F(t)|\Phi_a(t)
\rangle~.
\end{equation}
Comparing this expression with Eq.(\ref{ati}) we see that 
$A^{(e)}_\sigma$ are identical with the two terms
in the amplitude of the single-electron ionization, see Eq.(\ref{atij}).
The momenta
${\bf Q}_{\sigma,l}$ are defined in accordance
with Eqs.(\ref{R0tt}),(\ref{KFF}),(\ref{R0g0})
\begin{equation}\label{Qsig}
{\bf Q}_{\sigma,l}=(-1)^{\sigma}K_{l} \frac{ {\bf F} }{F},~~~\sigma=
1,2~.
\end{equation}
The quantity $R_{l}$ in Eq.(\ref{Afactsum}) is equal to 
\begin{equation}\label{R12abs}
R_{l}=\frac{F}{\omega^2}\sqrt{1-(\beta_{l}+i\gamma)^2},~~~~~~~~ 
\beta_{l}=
\frac{K_{l}\omega}{F}~.
\end{equation}
Evaluating this expression  one can suppose that the non-exponential factor
$R_0(t,t^{'})$ in Eq.(\ref{facrr}) is
$R_0(t,t^{'})\approx (-1)^{\sigma+1} 
(F/\omega^2) \cos x_\sigma,~\sigma=1,2$, see Eqs.(\ref{cos0}),
(\ref{R0g0}).
The quantity 
$\cos x_\sigma$ is to be found from Eq.(\ref{uvpkap}) in which
$p_{||}=(-1)^\sigma K_{l},~p_{\perp}=0$, and the sign in the right-hand
side  is $\pm 1=(-1)^{\sigma+1}$ in accordance with Eq.(\ref{cosg0}). 
As a result  the non-exponential factor
is  the same for both saddle points:
$R_0(t=x_1/\omega,t^{'})=
R_0(t=x_2/\omega,t^{'})=R_{l}$;  Re $R_{l}>0$, Im $R_{l}<0$. 

It follows from the definition of $K_{l}$ (\ref{Kmnm}) that
both amplitudes $A^{(e)}_1$ and $A^{(e)}_2$ in Eq.(\ref{Afactsum}) 
describe the ionization process with absorption of $l$ quanta.
Note however, that in the considered 
two-electron problem they manifest themselves
differently from the single-electron problem
see Eq.(\ref{atij}).  First, each
 of them depends on  its own argument, which is 
  the photoelectron momentum. It is ${\bf Q}_{\sigma,l}$ 
for $A^{(e)}_\sigma $, ${\bf Q}_{1,l}=-{\bf Q}_{2,l}$.
Second, they appear
in Eq.(\ref{Afactsums}) with different coefficients. 

The second factor in Eq.(\ref{Afactsum}) is
\begin{equation}\label{Ae2e}
A^{(e-2e)}(n-l; {\bf Q}_{\sigma,l};{\bf p},c)=-\frac{1}{2\pi}\frac{1}{T}
\int_{0}^{T}
\langle \Phi_{{\bf p}}(t)|V_{cb}(t)| \Phi_{{\bf Q}_{\sigma,l}}
(t)\rangle  dt~.
\end{equation}
It can be recognized as the amplitude of electron impact on the 
single-charged ion in the presence of the laser field, see Eq.(\ref{inppc}) in 
the next section. 
The initial and final momenta
of the electron are ${\bf Q}_{\sigma,l}$ and ${\bf p}$.
The collision  results in the excitation of the ion into the state $c$ 
and absorption of $n-l$ quanta.
The inequality (\ref{fom}) provides  the following restriction on $l$
\begin{equation}\label{resm}
l \ge l_0=\left [ \frac{F^2}{2\omega^3} \left( \gamma^2 +
\frac{1}{2} \right )\right ]_{>}~,
\end{equation}
where $[z]_{>}$ is a minimal natural number greater or equal to $z$, 
$z\le[z]_{>}$.
Eq.(\ref{resm}) has  a clear physical meaning: after absorption
of $l$ quanta the photoelectron must be in the continuum
state. Otherwise, if this inequality is broken,
the amplitude would be suppressed.
There will appear the  additional suppressing
 factor $\exp \{-\sqrt{(l_0-l)\omega}F/\omega^2 \}$ in the amplitude
as it follows from  Eq.(\ref{facrr}).
This  factor is governed by
the parameter
\begin{equation}\label{F32}
\frac{F}{\omega^{3/2}}>1~.
\end{equation}
The result of this section is the following representation for the
amplitude of two-electron process
\begin{equation}\label{smsig}
A^{(2e)}(n; {\bf p},c) = \sum_{l\ge l_0 } \frac{1}{R_{l}}\sum_{\sigma =1,2}
A^{(2e)}_\sigma(n,l; {\bf p},c)~.
\end{equation}
Here every term under summation over $l$ describes the 
process when first $l$ photons are absorbed during the single-electron
photo-ionization process and
then $n-l$ quanta are absorbed during
inelastic collision of photo-electron with the ion.
The term with $l=l_0$ describes the process which starts
 from the near-threshold
single-electron ionization. The other terms with $l>l_0$
describe the processes when  above-threshold single-electron ionization
takes place. It will be shown in Section V that
the role of the terms with $l=l_0$ and those with 
$l>l_0$ depends on the strength of
the field: for a sufficiently strong field the term with $l=l_0$ dominates, 
while for a weak field the ATI processes become extremely
important.

Every amplitude 
$A^{(2e)}_\sigma(n,l; {\bf p},c)$ in Eq.(\ref{smsig}) is presented
in the factorization form: it is the  product
of the amplitudes of photo-ionization and inelastic
collision, see Eq.(\ref{Afactsum}). Due to  largeness of the
 phases of the amplitudes $A^{(e)}_\sigma(n;{\bf Q}_{\sigma,l})$  it is 
 more convenient to take into consideration 
only one of them evaluating the other  
from the parity conservation law Eq.(\ref{AeP}). Then Eq.(\ref{smsig})
may be presented as
\begin{eqnarray}\label{Ae2eP}
\sum_{\sigma =1,2}&&
A^{(2e)}_\sigma(n,l; {\bf p},c)=\\ \nonumber
&& A^{(e)}_1(l; {\bf Q}_{l},c)
\left [ A^{(e-2e)}(n-l; {\bf Q}_{l};{\bf p},c)
+(-1)^{l}P_a A^{(e-2e)}(n-l; -{\bf Q}_{l};{\bf p},c)\right ]=\\ \nonumber
&&A^{(e)}_1(l; {\bf Q}_{l},c)\left [ A^{(e-2e)}(n-l; {\bf Q}_{l};{\bf p},c)
+(-1)^n P_a P_b A^{(e-2e)}(n-l; {\bf Q}_{l};-{\bf p},P(c))\right ]~.
\end{eqnarray}
Here ${\bf Q}_{l}={\bf Q}_{1,l}=-K_{l}{\bf F}/F$, and $P_b$
is the parity of the ion state  $b$. The last line
in Eq.(\ref{Ae2eP}) is evaluated taking account of Eq.(\ref{inP})
which presents the parity conservation law for the inelastic scattering 
amplitude discussed in the next section. 
Eqs.(\ref{smsig}),(\ref{Ae2eP}) are valid 
when the amplitude of the electron wiggling in the laser field is large, see
Eqs.(\ref{fom}),(\ref{abbrr}).

\section{Inelastic scattering}
Consider the inelastic electron-ion impact in the laser filed
restricting attention to the aspects of the problem which
are most closely related to the many-electron excitations
of the atom by the laser field.
The general expression for the amplitude of inelastic collision
in the first order of perturbation theory over  
electron-ion interaction is
\begin{equation}\label{inel}
A^{(e-2e)}=-\frac{1}{2\pi}\frac{1}{T}\int_{0}^{T}
\langle \Psi_f (t) | \frac{1}{r_{12}} | \Psi_i (t)\rangle ~,
\end{equation}
where $ \Psi_i (t), \Psi_f (t)$ are the initial and
final-state wave functions.
Note that Eq.(\ref{inel}) differs from Eq.(\ref{afi})
describing the general process.
In Eq.(\ref{inel})  both $\Psi_i (t)$ and $\Psi_f (t)$
account for  the interaction
of the system
with the laser field, while in Eq.(\ref{afi}) only the wave function
$\Psi_f (t)$ takes this interaction into account.
  The difference results from
the fact that  Eq.(\ref{inel})
from the very beginning describes 
the perturbation theory
over $1/r_{12}$. Note also that the factor $-1/(2\pi)$ in 
Eq.(\ref{inel}) is the usual one for a collision problem.

The wave functions for the  collision problem
should be chosen as
\begin{eqnarray}\label{psiininel}
&&|\Psi_i (t)\rangle=\Phi_{{\bf p}}({\bf r}_1,t)
\psi_b({\bf r}_2)\exp \{-i E_b t\}~,\\ \nonumber
&&|\Psi_f (t)\rangle=\Phi_{ {\bf p}^{'} }({\bf r}_1,t)
\Psi_c({\bf r}_2,t)~.
\end{eqnarray}
Here ${\bf p},{\bf p}^{'}$ are 
the momenta of the electron before and after
collision and $\psi_b,~\Psi_c$ are the wave functions 
of the ground and excited states of the ion. 
Substituting
Eq.(\ref{psiininel}) in (\ref{inel}) we find
\begin{eqnarray}\label{inppc}
A^{(e-2e)}(m,{\bf p};{\bf p}{'},c)=
&&-\frac{1}{2\pi}\frac{1}{T}
\int_{0}^{T}
\langle \Phi_{{\bf p}^{'}}(t)|V_{cb}(t)| \Phi_{\bf p}
(t)\rangle  dt=\\ \nonumber
&&-\frac{1}{2\pi}\frac{1}{T}\int_{0}^{T} 
\langle {\bf p}^{'},\phi_c (t) \left |\frac{1}{r_{12}} 
\right |{\bf p},\phi_b(t) \rangle
\exp \left \{\frac{i}{\omega}S_{{\rm sc}}(\omega t)\right \}dt ~,
\end{eqnarray}
where
\begin{equation} \label{Sinel}
S_{{\rm sc}}(x)=
\int_{0}^{x} \left [
\frac{1}{2}\left ({\bf p}^{'}+\frac{{\bf F}}{\omega} \sin x
\right )^2 +E_c(x/\omega)-\frac{1}{2}
\left ( {\bf p}+\frac{{\bf F}}{\omega} \sin
x \right )^2- E_b \right ] d x +\frac{({\bf p}- {\bf p}{'}){\bf F} }{\omega}~.
\end{equation}
Eq.(\ref{inppc}) demonstrates  that the previously considered in Eq.(\ref
{Ae2e}) quantity  is the amplitude of inelastic collision indeed.
The energy conservation law for the case under consideration reads 
\begin{equation}\label{inE}
\frac{  p^2}{2}+E_b+m\omega=\frac{  p^{'2}}{2}+\bar E_c~, 
\end{equation}
where $|m|$
is the number of  quanta which are absorbed, if $m>0$, or emitted, if $m<0$,
during the scattering.
The parity conservation law for the inelastic scattering reads
\begin{equation}\label{inP}
A^{(e-2e)}(m,-{\bf p};-{\bf p}{'},P(c))=
(-1)^m P_b A^{(e-2e)}(m,{\bf p};{\bf p}{'},c)~,
\end{equation}
where $P_b$ is the parity of the state $b$, and $P(c)$ is
the state with the quantum numbers  obtained by applying
the inversion operator $P$ to the quantum numbers of the state $c$.
It is easy to verify using Eqs.(\ref{PhiP}),(\ref{cP}) that the amplitude
Eq.(\ref{inppc}) satisfies Eq.(\ref{inP}).

As is usual in adiabatic processes there appears the large phase $\sim 1/
\omega$
in the integrand  in Eq.(\ref{inppc}). 
Therefore the saddle points give the major
contribution to $t$-integration. They  satisfy  the condition
\begin{equation}\label{Sprin}
 S^{'}_{{\rm sc}}(\omega t)=
\frac{1}{2}\left ({\bf p}^{'}+\frac{{\bf F}}{\omega} \sin \omega t
\right )^2 +E_c(t)-\frac{1}{2}
\left ( {\bf p}+\frac{{\bf F}}{\omega} \sin \omega t \right )^2- E_b=0~.
\end{equation}
It is fulfilled for those moments of $t$ when
the time-dependent ``energy'' of the system in the initial state coincides
with the ``energy'' in  the final state. If there is a real  saddle point 
$t$  then the process proceeds with a high probability. If it is 
in the complex plane, Im $t>0$, then 
there is a strong suppression originated from the adiabatic nature of the 
process. This statement can be checked out for special cases when
$t$-integration in Eq.(\ref{inppc}) can be fulfilled analytically.
One of them is the excitation into the discrete 
ion state $c$ having small polarizability. Then one can neglect 
the time-dependence of the energy
$E_c(t)$ and the corresponding wave function $\phi_c(t)$, 
$E_c(t)\approx E_c,~ \phi_c(t)\approx \phi_c$ and find
from Eq.(\ref{inppc}) the simple representation
\begin{eqnarray}\label{AfJ}
&&A^{(e-2e)}(m,{\bf p};{\bf p}{'},c)= f({\bf p};{\bf p}{'},c)J_m(
{\bf q F}/\omega^2) i^m,\\ \label{fppc}
&&f({\bf p};{\bf p}{'},c)=-\frac{1}{2\pi}
\langle {\bf p}^{'},\phi_c \left |\frac{1}{r_{12}} \right |{\bf p},\phi_b 
\rangle~,
\end{eqnarray}
where $f({\bf p};{\bf p}{'},c)$ is the scattering amplitude
in the absence of the laser field (considered outside the 
mass-shell),
$J_m(z)$ is a Bessel function and ${\bf q}={\bf p}-{\bf p}^{'}$
is the transferred momentum. 
Representation similar to Eq.(\ref{AfJ}) is well-known for the case of elastic
scattering considered in (Bunkin and Fedorov 1966).
A Bessel function is large
when its argument is greater than its index. Thus the amplitude Eq.(\ref{AfJ})
is not suppressed if $|{\bf q F}|/\omega^2\ge |m|$.
The latter condition is equivalent to one given in Eq.(\ref{Sprin})
for the considered case $E_c(t)=E_c$.

Consider now how much energy can be transferred to the ion
 from the electron and the field during collision.
Suppose that the initial-state
momentum ${\bf p}$ is given. The  final-state moment ${\bf p}^{'}$
will be considered as a free parameter and we will choose it to make
the energy level $E_c$ as high as possible.
From Eqs.(\ref{inE}),(\ref{Sprin}) we find that to make the process probable
the following inequality
is to be fulfilled  for some 
 real moment of time $t$ 
\begin{equation}\label{maxexc}
\frac{1}{2} ( {\bf p}+ {\bf k}_t )^2~
\ge E_c(t)-E_b~.
\end{equation}
The quantity in the left-hand side of Eq.(\ref{maxexc})
may be considered as a time-dependent ``energy'' of the impact electron.
To clarify the meaning of this equation 
consider first the  discrete-state excitations of the ion.
For the discrete state $c$ we can use Eq.(\ref{Ent}) to estimate its  energy. 
Then from Eq.(\ref{maxexc}) we find the maximum energy gap between the 
ion ground state $b$ and its excited  state $c$ which can be populated
in the collision 
\begin{equation}\label{EcEb}
{\rm max}(E_c-E_b)=\frac{1}{2} \left (| p_{||}| +\frac{F}{\omega}
\right )^2+\frac{1}{2}p_{\perp}^2 ~.
\end{equation}
Evaluating Eq.(\ref{EcEb})
it was assumed that $\alpha_c(\omega) \omega^2\le 1$.
Note that  in the left-hand side of Eq.(\ref{EcEb}) there stands
the non-shifted value of the energy $E_c$.
The moments of time $t$ for which the transition into the state $c$
satisfying Eq.(\ref{EcEb}) takes place are those for which the wiggling
energy is maximal 
\begin{equation}\label{sin0}
|\sin \omega t |=1~.
\end{equation}
Then we find that  final-state momentum ${\bf p}^{'}$ is to be
\begin{equation}\label{pfin}
|p^{'}_{||}|\approx \frac{F}{\omega},~~~ p_{\perp}{'}\approx 0~.
\end{equation}
Consider now the excitations of the ion into
continuum state, i.e. creation of a doubly charged ion. 
For this case $E_c(t) = ({\bf p}_1+{\bf k}_t)^2/2$, where ${\bf p}_1$
is the momentum of the knock-out electron. 
It is interesting to find the
maximum binding energy of the ion $E_b$ which would still allow the
collision to result in double ionization.
From Eq.(\ref{maxexc}) we find
\begin{equation}\label{mion}
{\rm max}|E_b| =\frac{1}{2} \left ( p_{||} +\frac{F}{\omega}
\right )^2+\frac{1}{2}p_{\perp}^2 ~,
\end{equation}
which agrees with Eq.(\ref{EcEb}) if we put in the latter one $E_c=0$.
Conditions (\ref{sin0}),(\ref{pfin}) are valid for this case as well. 
Eq.(\ref{sin0})
was used previously to evaluate the quantity $R_l$ 
in Eq.(\ref{R12abs}).

Note that according to Eq.(\ref {EcEb}) the maximum possible excitation
energy of the ion is equal to the  maximum of the ``energy''
of the impact electron. The latter is always higher,
and can be much higher,  than the averaged energy which stands in the 
energy conservation law: max$({\bf p}+{\bf k}_t)^2/2>\bar E_{\bf p}$.
This fact will be important in the following estimation of the probability of 
many-electron processes given in Section VI.

\section{qualitative discussion}
In the previous sections we considered  the collective response of two
atomic electrons to the laser field and found that the amplitude of the 
process may be presented as a product of the amplitude of single-electron 
ionization and the amplitude of inelastic scattering, see 
Eq.(\ref{Afactsum}) or (\ref{Ae2eP}). 
This means that at the beginning of the process single-electron ionization
takes place.
After that  the ionized electron undergoes inelastic collision  with
the parent atomic particle.
The general reason for factorization
of the amplitude is quite clear.  It follows from the existence of
two different scales  for electron localization.
One is the radius of the atom. The other is the radius of the wiggling of 
the electron in the laser field which is supposed to greatly  exceed the 
atomic radius.  

The simple physical meaning of the mathematical description
presented above is the following.
The single-electron ionization may be considered as a quasiclassical
movement of the atomic electron under the non-stationary, time-dependent
barrier. When the tunneling is finished the electron appears
from under the barrier at some particular point separated from the atom.
After that the electron
propagates  in a classically
allowed region. It strongly interacts with the laser field
exhibiting the wiggling. The propagation of the electron wave function
in the  classically allowed region
may proceed in any direction because it starts from the point. 
In particular it propagates
in the direction of the parent atomic 
particle. This makes possible the return of the electron
to the  single-charged ion.
As a result the collision of this electron with the ion is very
probable, the probability being proportional to the cross section and inversely
proportional to the square of the distance which separates the electron
from the atom 
after finishing the under-barrier part of the  trajectory. 
This distance is equal
to the absolute value of $R_l$ given in Eq.(\ref{R12abs})
\begin{equation}\label{rad}
|R_{l}|=\frac{F}{\omega^2}
\left \{[ (\beta_{l}+1)^2+\gamma^2][ (\beta_{l}-1)^2+\gamma^2]
\right \}^{1/2} ;
~~~~~~\beta_{l}=\frac{K_{l}\omega}{F},~~~\gamma=\frac{\kappa \omega}{F}~.
\end{equation}
It depends on the amplitude of the electron wiggling
$F/\omega^2$ as well as on $\gamma$ and the parameter $\beta_l$
which is the ratio of the photoelectron momentum to the field
momentum. The  photoelectron momentum  depends on the number of  absorbed 
quanta, which makes $|R_{l}|$  depend on it as well. 
This separation $|R_l|$ can exceed
the amplitude $F/\omega^2$ of the wiggling.
For this case wiggling itself is not sufficient to prove
that  the electron can return to the parent atomic particle. That is why
the discussion of all mathematical aspects of the problem
given in the previous section is vital.

There are two paths for the under-barrier 
electron escape from the atom. Roughly speaking, one of them is along the
direction of the field ${\bf F}$ and the other is in the opposite
direction. (The trajectory is described by  
complex time and coordinates. Therefore, it is more accurate to say that the
real part of the trajectory is either along the field or  in the opposite 
direction.)
Every  term $A^{(e)}_\sigma, \sigma =1,2$ in the amplitude of
single-electron ionization, see Eq.(\ref{atij}),
describes the escape along one of these two paths.
If we consider 
the pure single-electron ionization problem then after the tunneling
the momentum of the electron should be  the same for both routes of the escape.
In contrast, for the problem of  two-electron ionization the
electron must return to the parent atomic particle and have the collision with
it. The return route
is in the direction opposite to the escape path, being different
for  the events described by $ A^{(e)}_1$ and for those described 
by $ A^{(e)}_2$. 
This  explains why in Eq.(\ref{Afactsum}) the momentum  
${\bf Q}_{1,l}$, which is the argument of  $ A^{(e)}_1$, is  opposite 
to the field direction while the momentum ${\bf Q}_{2,l}$,
which is the argument of $ A^{(e)}_2$, is along this direction.

\section{estimation of the probability of many-electron process}

There are different ways for the process of  two-electron excitation
with absorption of $n$ quanta. 
 During ionization of the first electron
there may be absorbed some number 
$l$ of quanta.
The remaining number of quanta $n-l$ is absorbed
in the collision.
The two-electron amplitude is the sum of the amplitudes of these
independent events, see Eq.(\ref{smsig}).
The interference of the terms with
different $l$ should be  suppressed by the large and field-dependent phase
of the amplitude $A^{(e)}_1$, see discussion in Section II A.
Therefore as a first approximation we can neglect this interference. 
Then we find from Eq.(\ref{smsig}) the simple estimation for
 the probability $W^{(2e)}(c)$ of the two-electron process
when one electron in the final state is in the continuum and 
the other  is excited into the state $c$ belonging
to  the discrete or continuum  spectrum
\begin{eqnarray}\label{probn}
&&W^{(2e)}(c)=\sum_{n} W^{(2e)}(n;p_n,c),\\ \label{prob}
&&W^{(2e)}(n;p_n,c) = \sum_{l\ge l_{0}} 
\frac{d W^{(e)}}{d \Omega}(l,\theta=0)\frac{\sigma (n-l;K_{l};p_n,c)}
{|R_{l}|^2}~.
\end{eqnarray}
Here the total probability $W^{(2e)}(c)$
is presented as a sum of the probabilities
with absorption of a given number $n$ of quanta $W^{(2e)}(n;p_n,c)$.
The quantity  $d W^{(e)}(l,\theta=0)/d\Omega$
 is the differential probability of the single-electron
ionization with absorption of $l$ quanta,  and transition of the electron
into the state with momentum $K_{l}$ whose 
absolute value is  given in Eq.(\ref{Kmnm}), 
the condition $\theta=0$ shows  that the considered 
momentum direction is along the 
field, ${\bf K}_l=K_{l}{\bf F}/F$.
The quantity $\sigma (n-l;K_{l};p_n,c)$ in Eq.(\ref{prob}) 
is the cross section of the electron-ion
inelastic scattering in the presence of the laser field, $K_{l}$ and 
$p_n$ are the momenta
of the electron in the initial and final states for the scattering problem. 
The scattering
is accompanied by the absorption of $n-l$ quanta and   
excitation of the ion
into the state $c$. Evaluating Eq.(\ref{prob}) we took into 
account Eq.(\ref{prob2A}) for 
the probability of the single-electron event and for the sake of
simplicity neglected the interference of the two terms in the square
brackets in Eq.(\ref{Ae2eP}).
The distance  $|R_{l}|$ is given in Eq.(\ref{rad}).

Consider now how high is the excitation energy provided by
the considered mechanism.
The general rules showing that the excitation is probable
are  given by Eqs.(\ref{EcEb}),(\ref{mion}) in which  
 the impact electron momentum for the considered case is 
$|p_{||}|=K_l,~p_{\perp}=0$. As a result we come to Eq.(1)
\begin{equation}\label{aticex}
\frac{1}{2}\left( K_{l}+\frac{F}{\omega}\right )^2 \ge E_{\rm exc}~,
\end{equation}
which it is convenient to present as 
\begin{equation}\label{betggam}
\beta_l+1\ge \gamma_{\rm exc},~~~~~
\gamma_{\rm exc} = \frac{\omega \sqrt{2 E_{\rm exc}}}{F}~,
\end{equation}
where the quantity $\gamma_{\rm exc}$ may be called the adiabatic parameter for
the considered ion excitation. The energy $E_{\rm exc}$ is introduced to 
consider simultaneously the ion excitations into discrete and continuum
spectra: $E_{\rm exc}=E_c-E_b$ for discrete excitations and $E_{\rm exc}=
|E_b|$ for excitations into the continuum, 
compare Eqs.(\ref{EcEb}),(\ref{mion}).

Eq.(\ref{aticex})   demonstrates that the 
first ionized electron  must possess  ``enough energy'' to excite the ion. 
Due to wiggling in the laser field the ``energy'' of the electron 
is not conserved, it oscillates.   Condition (\ref{aticex})
states that  at the most favourable moment of time,
when the field momentum is big, $k_t=F/\omega$, and
parallel to the translational momentum $K_l$, 
this energy must be equal to the excitation energy.
Note that the averaged energy of the photoelectron
is always lower than the maximum one:
$(K_l+F/\omega)^2/2>\bar E_{{\bf K}_l}= K_l^2/2+ F^2/(4\omega^2)$.
For further consideration 
it is convenient to  distinguish the case  of  high wiggling energy
from the case of low wiggling energy.
Consider them in turn.\\
1. High wiggling energy.\\
Suppose that the energy of the electron wiggling is so high that it 
exceeds the ion excitation energy for the considered state $c$, 
$F^2/(2 \omega^2)\ge E_c-E_b$, i.e.
\begin{equation}\label{hw}
\gamma_{\rm exc} \le 1~.
\end{equation}
Then it is sufficient to take into account only one first
term $l=l_0$ in  the sum over $l$ in Eq.(\ref{prob}).
It describes the process when first
the near-threshold single-electron ionization takes place. 
The momentum of photoelectron is small $K_{l_0}\approx 0$.
The electron wiggling  provides enough  energy to excite the ion 
during  collision as is  seen  from the conditions 
(\ref{betggam}),(\ref{hw}).
The probability of the process may be estimated as 
\begin{equation}\label{probnt}
W^{(2e)}(c) \approx \frac{d W^{(e)}}{d \Omega}(l_0,\theta=0)
\frac{\sigma (n-l_0;K_{l_0};p_n,c)}{|R_{n}|^2}~.
\end{equation}
Eq.(\ref{rad}) for the considered case reduces to 
\begin{equation}\label{ntr}
|R_{l_0}|=\frac{F}{\omega^2}\sqrt{1+\gamma^2}~.
\end{equation}
2.Low wiggling energy. \\
Suppose that the wiggling energy is small compared with the excitation energy,
${F^2}/{2 \omega^2} < E_c-E_b$,
\begin{equation}\label{lw}
\gamma_{\rm exc} > 1~.
\end{equation} 
Then the wiggling  energy $F^2/(2\omega^2)$ is insufficient
to provide the ion excitation.
Therefore in this case we are to consider ATI during the single-electron 
ionization described by the  terms $l>l_0$ in the 
sum in Eq.(\ref{prob}).
Let us call $l_1$ the minimum  number of quanta the absorption 
of which satisfies
Eq.(\ref{aticex}). From Eqs.(\ref{aticex}),(\ref{Kmnm}) we find
\begin{equation}\label{l1l0}
l_1=\left [\frac{F^2}{2\omega^3}\left ((\gamma_{\rm exc}-1)^{2}
+\gamma^2+\frac{1}{2}
\right ) \right ]_{>}>l_0=
\left [ \frac{F^2}{2\omega^3}\left (\gamma^2+\frac{1}{2}\right ) \right ]_{>}~.
\end{equation}
The inequality (\ref{betggam}) shows that
for a very weak field when $\gamma_{\rm exc}>>1$,
the above threshold energy must
be equal to the ion excitation energy: $K_{l_1}^2/2\approx E_c-E_b$. 
In this case $l_1$ can substantially
exceed  $l_0$.
With increase of the field the necessary above threshold energy
rapidly decreases. Really, it is seen from Eq.(\ref{l1l0}) 
that if $\gamma_{\rm exc} \sim 1$, then $l_1\sim l_0$. In this case
the purpose of ATI is to supply the ionized electron with
the momentum $K_l$ which combined with the large field momentum 
will give the necessary high maximum ``energy'' for the photoelectron, see
Eq.(\ref{aticex}). At the
same time its averaged energy  may be 
much lower than the ion excitation energy.
The example of two-electron process in He considered in Section VII 
illustrates this possibility. We 
come to a conclusion that there are two reasons for increase of 
the probability of the two-electron process
with increase of the field
in the region Eq.(\ref{lw}).
First, the total probability of single-electron ionization grows. Second,
the stronger the field the lower becomes
the necessary level in the ATI  spectrum.

The estimation for  the  probability of 
two-electron process is
\begin{equation}\label{probnt1}
W^{(2e)}(c) \approx  \frac{d W^{(e)}}{d \Omega}(l_1,\theta=0)
\frac{\sigma (n-l_1;K_{l_1};p_n,c)}{|R_{l_1}|^2}~,
\end{equation}
where $R_{l_1}$ is given in Eq.(\ref{rad}).

The total number of absorbed quanta $n$ in the processes
described by Eqs.(\ref{probnt}),(\ref{probnt1})
must be large enough to provide the excitation of the level $c$ as well as
a sufficiently large final state momentum $p_n$ of the photoelectron:
$p_n\approx{F/\omega}$, see Eq.(\ref{pfin}).
Therefore,   the averaged energy of the scattered electron is
$\bar E_{ {\bf p}_n}
\approx 3 F^2/(4\omega^2)$. 
Estimation for the number of absorbed quanta $n$ follows from the
conservation law
\begin{equation}\label{estn}
n \approx n_{\rm min}+\left [\frac{F^2}{2\omega^3}\right ]_{>},~~~~~
n_{\rm min}=\left [ \frac{1}{\omega}\left ( 
\bar E_c-E_a-E_b+\frac{ F^2}{4 \omega^2} \right ) \right ]_{>}=
\left [\frac{F^2}{2\omega^3} \left (\gamma_{\rm exc}^2
+\gamma^2+\frac{1}{2}
 \right ) \right ]_{>}~.
\end{equation}

\section{double-ionization of He atom by 780 nm laser field}
In order to illustrate the ideas discussed in the previous section consider
the recent experimental results 
(Walker $et~al$ 1994) on precision 
measurement of double ionization of He
atom by the 780 nm laser field in the intensity region from 0.15 to 5.0
PW$/$cm$^2$. The data show the high ratio of He$^{++}$ to He$^{+}$ yield
in this region of  intensities. For the lowest 
intensity  0.15 
PW$/$cm$^2$ the field momentum is $F/\omega=1.2$ atomic units.
Therefore 
the maximal wiggling energy $F^2/(2 \omega^2)=19.3$ eV is well below
the lowest excitation energy of He$^+$ which is  41 eV.
Absorption of only 3 quanta above the single-electron ionization threshold
changes the situation drastically. 
It gives the translational
momentum $K=0.58$. As a result the 
maximal energy of photoelectron becomes $(K+F/\omega)^2/2=42.8$ ev.
The above consideration, see Eq.(\ref{aticex}) or (\ref{maxE}),
shows that this maximal energy of photoelectron can be used for ion
excitation. Thus we see that absorption of only 3 above-threshold quanta
during the single-electron ionization 
makes   the excitation of He$^{+}$ possible. The excitation might  result
in the double charged ion formation due to field ionization from
the excited state. 
Note that the above-threshold excitation energy itself in the 
considered example is sufficiently low. After absorption of 3 quanta 
it is only $\sim 5$ eV which is far below
the  excitation energy. This illustrates  that 
the role of ATI is to supply the ionized electron with the momentum, rather
than with above-threshold energy in ATI spectrum. 
Similarly, the absorption of 6 above-threshold quanta  makes
possible the direct excitation of the ion into the continuum, i.e. 
double-ionization.
The probability of absorbing several quanta above
the threshold is high for the considered region of intensities of the laser 
field, see for example the recent work 
(Schumacher $et~al$ 1994)
where ATI
for the first and second harmonics of 1064 nm laser field with total 
intensity of
$4\times 10^{13}$ and $8\times 10^13$ W$/$cm$^2$ was examined. 
Thus the proposed mechanism  qualitatively agrees with the 
experimental results.

\section{conclusion}
We considered the mechanism of the many-electron process
when first single-electron ionization takes place and then the ionized electron
absorbs energy from the laser field and transfers it to the parent
atomic particle. We proved by explicit calculations that this
mechanism is very probable and provides the high energy necessary for 
ion excitation. The amplitude
of the two-electron process is presented as a product
of the amplitude of single-electron ionization  and the
amplitude of inelastic scattering. Both later quantities can
be calculated  $ab~ inicio$ as discussed in the Sections II A and IV. 
Therefore the presented results provide a possibility for accurate 
calculations of many-electron processes in the laser field. 
The simple estimations for the probability of the
process are also possible as discussed in Sections VI,VII.
ATI proves to be very important for the mechanism considered. It supplies
the ionised electron with an additional translational momentum. The latter
gives sharp rise to the maximal energy of the photoelectron resulting
in the possibility of excitation of the high-energy levels of the ion.

The results obtained can be developed  to cover 
related phenomena. One of them is  ATI itself. We have seen that
the scattering of the ionised electron on the parent atomic particle 
results in the transformation of its maximum  energy  into
the excitation of the ion. Another possibility is to transform
the maximum total energy into the energy of translational movement.
This would strongly increase the probability of the population of 
high-energy ATI levels. 
For the first time this idea was discussed in (Kuchiev 1987). 
Another related problem is the high-energy gamma-quanta production by 
an atom in a  laser field. 

In this paper we described the continuum states with the help of
the Volkov functions. This description needs improvement, as the static
Coulomb field in some cases is important, see for example 
(Reiss and Krainov 1994).
One can expect this improvement to  modify the amplitude of
single-electron ionization and the scattering amplitude. It can also
modify the parameter $R_l$ describing the separation between
the ion and first ionised electron. However,
factorization as a general property should remain valid. 
The reason for
this comes from the fact that factorization is based on the parameter which is
the large amplitude of classical wiggling of the electron in the laser field.
For a sufficiently strong field this quantity can remain large enough even
if one takes into account the ion Coulomb field.

The other approximation made in the paper is the consideration of only the
first order of perturbation theory over the electron interaction. 
This approximation can be insufficient if we are interested in the absorption
of the very large number of quanta. The reason for this is 
that the electron impact transforms electron energy into the ion excitation,
but there is always a possibility for  the scattered electron to absorb more 
energy from the field and again transfer it to the ion during one more 
collision with  it.

\acknowledgments
It is my pleasure  to acknowledge V.V.Flambaum,
G.F.Gribakin and O.P.Sushkov for valuable discussions. The kind help
of L.S.Kuchieva in preparation of the manuscript is appreciated.
The financial support of the Gordon Godfrey Fund of the University
of New South Wales and the Australian Research Council
is acknowledged.


\newpage
\begin{references}

Agostini P, Fabre F, Maifray G, Petit G, and Rahman N 1979
Phys.Rev.Lett.{\bf 42}, 1127\\

Agostini P and Petite G 1984
J.Phys.B:At.Mol.Phys.{\bf 17}, L811\\


Agostini P and Petite G 1985a
J.Phys.B:At.Mol.Phys.{\bf 18}, L281\\


Agostini P and Petite G 1985b
Phys.Rev.A {\bf 32}, 3800\\


Bondar' I I and Suran V V 1993
JETP {\bf 76}, 381\\

Boyer K, Egger H, Luk T S, Pummer H, and Rodes C K 1984
Opt.Soc.Am.B {\bf 1}, 3\\


Boyer K and Rhodes C K 1985
Phys.Rev.Lett.{\bf 54}, 1490\\ 

Bunkin F V and Fedorov M V 1966 Sov.Phys.JETP, {\bf 22}, 844\\

Camus P, Kompitsas M, Cohen S, Nicolaides C,
Aymar M, Crance M, and Pillet P 1989
J.Phys.B:At.Mol.Opt.Phys.{\bf 22}, 445\\

Chin S L, Yergeau F, and Lavigne P 1985
J.Phys.B:At.Mol.Phys.{\bf 18}, L213\\

Corkum P B 1993 Phys.Rev.Lett. {\bf 71},1994\\


Dexter J L, Jaffe S M, and Galagher T F 1985
J.Phys.B:At.Mol.Phys.{\bf 18}, L735\\

DiMauro L F, Kim D, Courtney M W, and Anselment M 1988
Phys.Rev.A {\bf 38}, 2338\\

Eichmann U, Zhu Y, and Gallagher T F 1987
J.Phys.B:At.Mol.Phys.{\bf 20}, 4461\\


Feldmann D, Krautwald H J, and Welge K H 1982a
J.Phys.B:At.Mol.Phys.{\bf 15}, L529\\


Feldmann D, Krautwald H J, Chin S L, von Hellfeld A, and Welge K H 1982b
J.Phys.B:At.Mol.Phys.{\bf 15}, 1663\\

Fittinghoff D N, Bolton P R, Chang B, and Kulander K C 1992
Phys.Rev.Lett. {\bf 69}, 2642\\

Freeman R R and Bucksbaum P H 1991
J.Phys.B:At.Mol.Opt.Phys.{\bf 24}, L325\\



Johann U, Luk T S, Egger H, and Rodes C K 1986
Phys.Rev.A {\bf 34}, 1084\\


Keldysh L V 1965 Sov.Phys.JETP {\bf 20}, 1307\\


Kuchiev M Yu 1987 Sov.Phys.JETP Lett. {\bf 45}, 404\\


Lambropoulos P 1985
Phys.Rev.Lett.{\bf 55}, 2141\\


L'Huillier A, Lompre L A, Mainfray G, and Manus C 1982
Phys.Rev.Lett.{\bf 48}, 1814\\


L'Huillier A, Lompre L A, Mainfray G, and Manus C 1983a
Phys.Rev.A {\bf 27}, 2503\\

L'Huillier A, Lompre L A, Mainfray G, and Manus C 1983b
J.Phys.B:At.Mol.Phys.{\bf 16}, 1363\\


Lompre L A, L'Huillier A,  Mainfray G, and Fan J Y 1984
J.Phys.B:At.Mol.Phys.{\bf 17}, L817\\


Luk T S, Pummer H, Boyer K, Shahidi M, Egger H, and Rhodes C K 1983
Phys.Rev.Lett.{\bf 51}, 110\\


Ostrovsky V N, Telnov D A 1987a J.Phys.B:At.Mol.Phys.{\bf 20}, 2397\\

Ostrovsky V N, Telnov D A 1987b
J.Phys.B:At.Mol.Phys.{\bf 20}, 2421\\

Potvliege R M and Robin Shakeshaft 1989
Phys.Rev.A {\bf 40} 3061\\



Reiss H R 1980 Phys.Rev.A {\bf 22}, 1786\\


Reiss H R 1987 J.Phys.B:At.Mol.Opt.Phys.{\bf 20}, L79\\


Reiss H R  and Krainov V P 1994
Phys.Rev.A {\bf 50} R910\\


Schumacher D W, Weihe F, Muller H G, and Bucksbaum P H 1994
Phys.Rev.Lett.{\bf 73}, 1344\\



Sz$\ddot o$ke A and Rhodes C K 1986
Phys.Rev.Lett.{\bf 56}, 720\\

Suran V V, and Zapesochnyi' I P 1975 Sov.Tech.Phys.Lett.,
{\bf 1}, 420\\



Tang X and Lambropoulos P 1987
Phys.Rev.Lett.{\bf 58}, 108\\


Volkov D M 1935 Z.Phys. {\bf 94}, 250\\


Walker B, Sheehy B, DiMauro L F, Agostini P, Shafer K L, and 
Kulander K C 1994
Phys.Rev.Lett.{\bf 73}, 1227\\



Zhu Y, Jones R R, Sandner W, Gallagher T F,
Camus P, Pillet P, and Boulmer J 1989
J.Phys.B:At.Mol.Opt.Phys.{\bf 22}, 585\\

\end{references}
\end{document}